# "An Approach for Controlling Faults in Wireless Sensor Networks Using Clustering"

Touseef Yousuf[1], Aminuddin Zabi[2] and Pallavi M[3].

*[1,2]M.TECH Scholar at NMIT, Bangalore India.*

*[3]Assistant professor, CS&E, NMIT, Bangalore, India.*

*Abstract*—Fault control and tolerance in WSN is a challenging problem because of limited energy, bandwidth, and computational complexity. While facing numerous threats these severely resource constrained nodes are responsible for data collection, data processing, localization, time synchronization aggregation and data forwarding. One of the effective approaches to control and tolerate these threats is through clustering.  In this paper we present a new method called EFCM "Efficient Fault Control Mechanism" for fault controlling in wireless sensor networks based on clustering and cluster-head selection. Simulation results show EFCM has better performance over state of art methods.

*Keywords*—WSN, X-means, Round-robin, Clustering, Fault Tolerance, EFCM, Cluster Construction, Cluster-head selection.

## I. Introduction

Wireless sensor networks (WSNs) are composed of numerous small-size sensors that have various features, such as low cost, light weight, high mobility, and capabilities for sensing, computing, and communications. The sensor has the same capabilities as the machine does. In fact, WSNs are one of typical application of machine to machine communications..

   They usually monitor areas, collect data and report to the base station (BS). Due to the achievement in low power digital circuit and wireless communication, many applications of the WSN are developed and already been used in habitat monitoring, military object and object tracking [1]. The energy being limited resource, consumption can be reduced by allowing only a portion of the nodes, which are called cluster heads to communicate with the base station.

    After the clusters are constructed and selection of cluster heads is completed the cluster heads communicate data with base station [2]. These in-efficient cluster heads cannot maximize energy efficiency.

    In this paper, we devised a new distributed clustering approach called EFCM which stands for "Efficient Fault Control Mechanism" that considers energy and time. The cluster heads are selected periodically after a particular time-span based on round-robin Scheduling and commensurate with the energies sorted in descending order. The remainder of this paper is structured as follows.

Section 2 briefly describes the related work. Section 3 gives an overview of taxonomy. Section 4 presents our working principle. Section 5 presents the simulation environment and experimental results. Finally conclusions are drawn in Section 6.

## II. Related work

In this section, we briefly review the related works in the area of fault detection in wireless sensor networks (WSNs). The existence of faulty sensor measurements in WSNs will not only cause degradation of the quality of service, but also lays a huge burden on the constraint energy of each sensor node. Several papers have proposed fault-tolerant event detection techniques. Failures in wireless sensor networks can occur for various reasons. First, sensor nodes are fragile, and they may fail due to depletion of batteries or destruction by an external event. In addition, nodes may capture and communicate incorrect readings because of environmental influence on their sensing components. Second, as in any ad hoc wireless networks, links are failure-prone [3], causing network partitions and dynamic changes in network topology.

    Links may fail when permanently or temporarily blocked by an external object or environmental condition. Packets may be corrupted due to the erroneous nature of communication. In addition, when nodes are embedded or carried by mobile objects, nodes can be taken out of the range of communication. Third, congestion may lead to packet loss. Congestion may occur due to a large number of nodes simultaneous transition from a power saving state to an active transmission state in response to an event-of-interest [4].

    Furthermore, all of the above fault scenarios are worsened by the multi-hop communication nature of sensor networks. It often takes several hops to deliver data from a node to the sink; therefore, failure of a single node or link may lead to missing reports from the entire region of the sensor network. Additionally, congestion that starts in one local area can propagate all the way to the sink and affect data delivery from other regions of the network. One of the approaches is clustering. Wireless sensors in a network are divided into different virtual groups, and they are allocated geographically adjacent into the same cluster according to

some rules with different behaviors for nodes included in a cluster from those excluded from the cluster. A typical cluster structure is shown in Fig. 1. It can be seen that the nodes are divided into a number of virtual groups based on certain rules. Under a cluster structure, mobile nodes may be assigned a different status or function, such as cluster-head, cluster gateway, or cluster member. A cluster-head normally serves as a local coordinator for its cluster, performing intra-cluster transmission arrangement, data forwarding, and so on. A cluster gateway is a non-cluster-head node with inter-cluster links, so it can access neighboring clusters and forward information between clusters. A cluster member is usually called an ordinary node, which is a non-cluster-head node without any inter-cluster links.

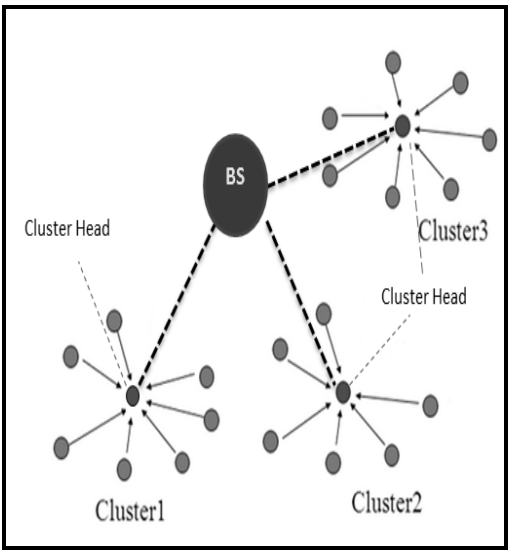

Fig. 1. Architecture

**A. Need for Clustering**

A cluster structure, as an effective fault control means [5], provides at least three benefits. First, a cluster structure facilitates the spatial reuse of resources to increase the system capacity [6]. With the non-overlapping multi cluster structure, two clusters may deploy the same frequency or code set if they are not neighboring clusters. Also, a cluster can better coordinate its transmission events with the help of a special mobile node such as a cluster head residing in it.This can save much resources used for retransmission resulting from reduced transmission collision. The second benefit is in routing, because the set of cluster-heads and cluster gateways can normally form a virtual backbone for inter-cluster routing, and thus the generation and spreading of routing information can be restricted in this set of nodes. Last, a cluster structure makes an ad hoc network appear in the view of each mobile terminal. When a mobile node changes its attaching cluster, only mobile nodes residing in the corresponding clusters need to update the information [7]. Thus, local changes need not be seen and updated by the entire network, and information processed and stored by each mobile node is greatly reduced.

### III. Taxonomy of Fault-tolerance

Recent research has developed several techniques that deal with different types of faults at different layers of the network stack. To assist in understanding the assumptions, focus, and intuitions behind the design and development of these techniques, we borrow the taxonomy of different fault tolerant techniques used in traditional distributed systems.

Fault prevention is to avoid or prevent faults. Fault detection is to use different metrics to collect symptoms of possible faults. Fault isolation is to correlate different types of fault indications (alarms) received hypotheses from the network, and proposes various fault. Fault identification is to test each of the proposed hypotheses in order to precisely localize and identify faults. Fault recovery is to treat faults, i.e., reverse their adverse effects.

Clustering techniques for WSNs can be used to overcome the above flaws by generally classifying the overall network architectural and operation model and the objective of the node grouping process including the desired count and properties of the generated clusters. In this section we discuss the different classifications and present taxonomy of a clustering attributes.

*A. Network model*

Different architectures and design goals and constraints have been considered for various applications of WSNs. In the following literary text procure some the relevant architectural parameters and accentuate their implications on network clustering.

*1. Network Concepts:*

Basically WSNs consist of three main components, sensor nodes, base-station and monitored events. Aside from the few setups that utilize mobile sensors [8], most of the network architectures assume that sensor nodes are stationary. Sometimes it is deemed necessary to support the mobility of base-station or cluster heads. Node mobility would make clustering very challenging since the node membership will dynamically change, forcing clusters to evolve over time. On the other hand, the events monitored by a sensor can be either intermittent or continual depending on the application. For instance, in target detection or tracking application, the event is dynamic whereas forest monitoring for early fire prevention is an example of intermittent events. Monitoring intermittent events allows the network to work in a reactive mode, simply generating traffic when reporting. Continual events in most applications require periodic reporting and consequently generate significant traffic to be routed to the sink. Although continual events would mostly make the clusters stable, it may unevenly load cluster heads relative to the nodes in the cluster and a rotation of the cluster head role may be required if the cluster head is randomly picked from the sensor population. Intermittent events would favor adaptive clustering strategies if the number of events significantly fluctuates.

*2. Data processing:*

Data aggregation combines data from different sources by using functions such as suppression, eliminating duplicates, min, max and average. Some of these functions can be performed either partially or fully in each sensor node, by allowing sensor nodes to conduct in-network data reduction. Data aggregation is also feasible through signal processing techniques.

*3. Topological deployment of nodes*:

This is application dependent and affects the need and objective of the network clustering. The deployment is either deterministic or self-organizing. In deterministic situations, the sensors are manually placed and data is routed through pre-determined paths. Therefore, clustering is such setup is also preset or unnecessary. However in self-organizing systems, the sensor nodes are scattered randomly creating an infrastructure. In that infrastructure, the position of the base-station or the cluster head is also crucial in terms of energy efficiency and performance.

*B. Cluster Properties*

Often clustering schemes strive to achieve some characteristics for the generated clusters. Such characteristics can be related to the internal structure of the cluster or how it relates to others. The following are the relevant attributes:

*1. Cluster Count:*

In some published approaches the set of cluster heads are predetermined and thus the number of clusters are preset. Randomly picking cluster heads from the deployed sensors usually yields variable number of clusters.

*2. Stability:*

When the clusters count varies and the node's membership evolves overtime, the clustering scheme is said to be adaptive. Otherwise, it is considered fixed since sensors do not switch among clusters and the number of clusters stays the same throughout the network lifespan.

Intra-cluster topology clustering schemes are based on direct communication between a sensor and its designated cluster head. However, multi-hop sensor-to-cluster head connectivity is sometimes required; especially when the sensor's communication range is limited and/or the cluster head count is bounded.

*3. Inter-cluster head connectivity:*

When the cluster head does not have long haul communication capabilities, cluster heads connectivity to the base-station has to be provisioned. In that case, the clustering scheme has to ensure the feasibility of establishing an inter-cluster head route from every cluster head to the base-station. Some of the published work assumes that cluster head would be able to directly reach the base-station.

*C. Cluster head capabilities*

As discussed earlier the network model influences the clustering approach; particularly the node capabilities and the scope of the in-network processing. The following attributes of the cluster head node are differentiating factors among clustering schemes:

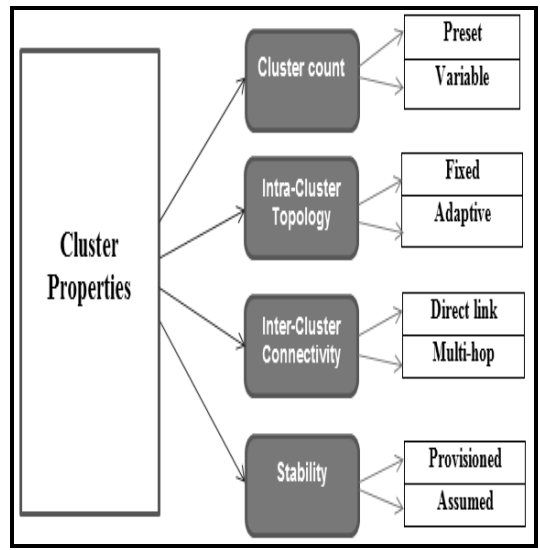

Fig. 2. Cluster properties

*1. Mobility:*

When a cluster head is mobile, sensor's membership dynamically changes and the clusters would need to be continuously maintained. On the other hand, stationary cluster head tends to yield stable clusters and facilitate intra- and inter-cluster network management. Sometimes, cluster heads can travel for limited distances to reposition itself for better network performance.

*2. Node Types:*

As indicated earlier, in some setups a subset of the deployed sensors are designated as cluster heads while in others cluster heads are equipped with significantly more computation and communication resources.

*3. Role:*

A cluster head can simply act as a relay for the traffic generated by the sensors in its cluster or perform aggregation/fusion of collected sensors' data. Sometime, a cluster head acts as a sink or a base-station that takes actions based on the detected phenomena or targets.

Congestion may lead to packet loss. Congestion may occur due to a large number of nodes simultaneous transitions from a power saving state to an active transmission.

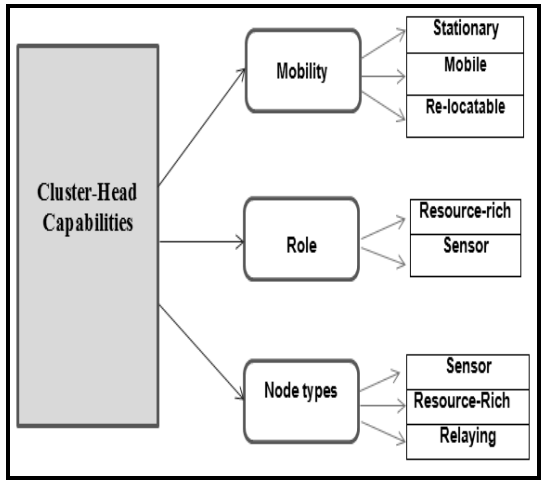

Fig. 3. Cluster-head capabilities.

*D. Clustering Proceedings*

The coordination of the entire clustering process and the characteristics of the algorithms vary significantly among published clustering schemes. The following attributes are deemed relevant [9]:

*1. Approach:*

When cluster heads are just regular sensors nodes, clustering has to be performed in a distributed manner without coordination. In few approaches, a centralized authority partitions the nodes offline and controls the cluster membership. Hybrid schemes can also be found; especially when cluster-heads are rich in resources. In the later case, inter-cluster-heads coordination is performed in a distributed manner, while each individual cluster head takes charge of forming its own cluster. Nodes assemble: As discussed in the previous section, several objectives have been pursued for forming clusters. Examples include fault-tolerance, load balancing, and network connectivity. Cluster-heads can be pre-assigned or picked randomly from the deployed set of nodes.

## IV. PROPOSED SCHEME

In this section we briefly describe the framework of our EFCM algorithm and its operating phases i.e. (1) Cluster construction phase and (2) Cluster-head selection phase.

*A. Frame Work*

The structure of the EFCM algorithm is chronicled in the succeeding steps as follows:

Step-1: We start with clustering the randomly deployed nodes using X-means [10,11] clustering method.

Step-2: The energy of all the nodes is obtained and sorted with respect to clusters [20].

Step-3: The cluster member which has the highest energy will be selected as a cluster-head for certain time-slice [19].

Step-4: After the cluster head is selected it announces its status to all its cluster members.

Step-5: When the set time-slice of cluster head expires, then the next cluster member which has the highest energy is selected as cluster head in a round-robin fashion.

Step-6: Selection of cluster head is done such that every cluster member of every cluster gets the opportunity to become a cluster head.

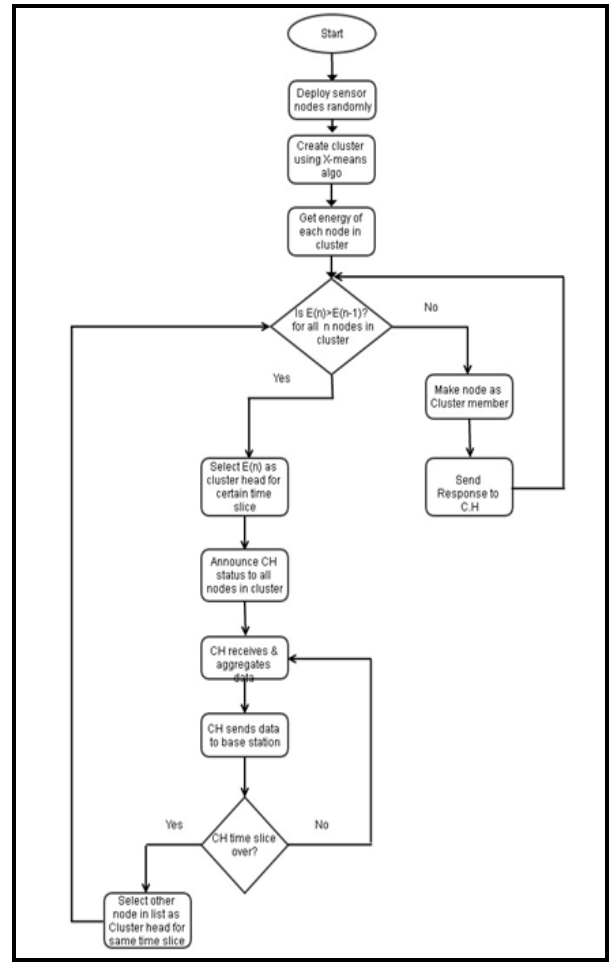

Fig. 5. Frame Work of EFCM Algorithm

*B. Cluster Construction*

In this phase, nodes in the network are clustered using X-means algorithm [10, 11] .We start constructing the clusters by initially setting the cluster list to NULL (Line 1-2), then we apply x- means to form clusters from the initial given set of nodes (Line 3-6). Once the clusters are formed we calculate the energy [20] of each node with respect to its cluster (Line 8-12), the energy values are sorted in decreasing order (Line 13), so that it can be used to select the initial cluster-head in each of the cluster in the cluster-head selection algorithm.

```
Cluster_Construction (v)
{
    (1) C (v) ← M (v); where M(v) ∈ N  // list of nodes for clustering
    (2) X (i) = NULL  // initial list of clusters is empty
    (3) For u from first to the last node
    (4)     X (i) ← X-Means(u) //clustering by X-means.
    (5)     Increment i by 1
    (6) End For
    (7) For j from 1 to i
    (8)   For n from first to last node in c(j)
    (9)     E(m) ← energy of n // energy of all cluster members in a cluster.
    (10)    Increment m by 1
    (11)  End for
    (12) End for
    (13) Sort E (m) in decreasing order
    (14) Cluster_Head_Selection()
}
```

Fig. 6. Cluster construction algorithm.

### C. Cluster-head Selection

In this phase a cluster head is elected for each cluster by taking into consideration of energy as well as time slice. Here we lay emphasis of selecting the cluster head in round robin fashion so that every node gets the opportunity of becoming a cluster head for set time slice. Initially in each cluster the node with the highest energy is selected as cluster-head(Line2-3).This node remains the cluster-head for set time slice(Line3-5).After time slice gets over other nodes in the cluster get the opportunity to become the cluster-head for set time slice in the round robin fashion.

```
Cluster-Head_Selection (time, E(m))
{
    (1) If time mod time-slice is zero
    (2) For k from 1 to m
    (3) if  E(k) > E(k+1)
    (4) E(k) ← Cluster-head  //for obtained time-slice
    (5)  End for
    (6) Announce the selected cluster head to all the cluster members of a particular cluster.
    (7) Endif
    (8) Goto step (1)
}
```

Fig. 7. Cluster-head selection algorithm.

## V. SIMULATION ENVIRONMENT

### A. Performance Metrics

We evaluate mainly the performance of *EFCM* algorithm according to the following metrics, by varying the pause time.

**1. Throughput:** It is the number of bits that are conveyed or processed per unit of time.

**2. Packet Delivery Ratio:** It is the ratio of the no. of packets received successfully and the total no. of packets sent.

**3. Residual Energy:** Specifies the energy possessed by a sensor node at a given point of time.

**4. Cluster-head failure:** This specifies the failure probability of each cluster-head in each round.

In Fig. 8 initially EFCM updates the routing table so as to establish path from source to base station (as shown in the when the time 0 to 5 sec).

However with the increase in time the throughput increases exponentially that ensures high data transmission from source to base station. Increase in throughput is due to multiple route capability of EFCM protocol (from 5 sec onwards).Further from above plot we can determine our *EFCM* protocol provides better throughput than LEACH and HEED.

In Fig.9 as the time increases the number of possible paths between source and sink pair increases which causes a significant increase in the exploratory messages within in network which initially leads to higher packet delivery ratio of about 95-100% in case of EFCM which is higher than LEACH(about 90%) and HEED(about75%). Even though initially the packet delivery ratio of the LEACH and HEED is higher than EFCM but as time increases there is a significant decrease in in packet delivery ratio in contrast to the EFCM where packet delivery ratio remains stabilized throughout. For our EFCM protocol, from above plot we can deduce that the energy dissipation is very slow as compared to LEACH and HEED (exponential decrease) thereby lifetime of network is increased. After 20 sec there is only 14-15% decrease in energy in EFCM where as in LEACH and HEED it's about 25 % and 23% decrease in energy.

dissipations of a node should be gradual over a period of time so as to increase the longevity of the network.

Cluster head failure should be minimal so as to achieve high throughput. As from above plot we can determine cluster head in EFCM protocol has greater lifespan as compared to LEACH and HEED represented in Fig. 11 protocol so there high throughput and less repair overhead of the network. Further in EFCM ensures a fair distribution of load among cluster heads which in turn increase the longevity of the network.

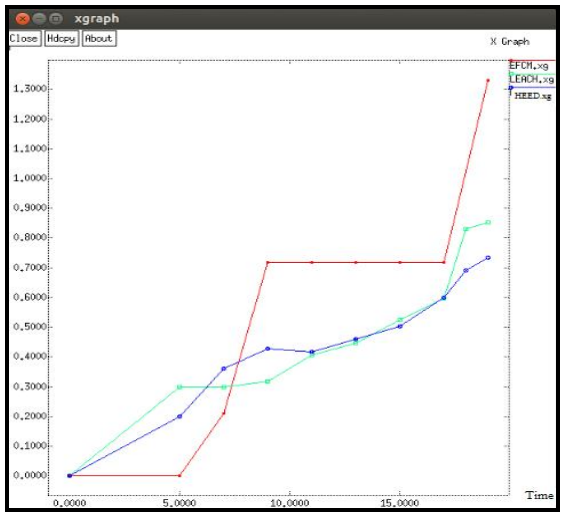

Fig. 8. Throughput vs. pause time.

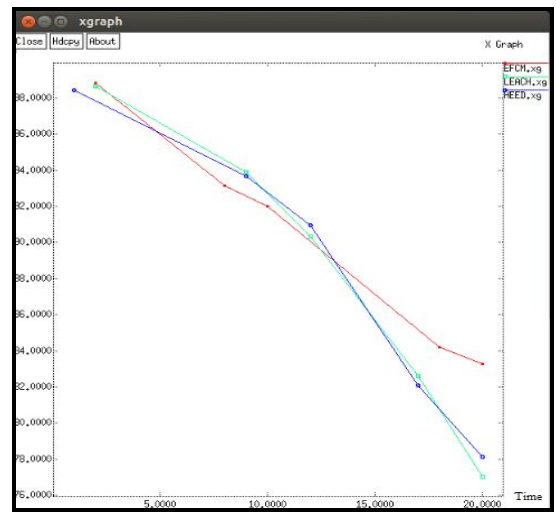

Fig. 10. Average residual energy.

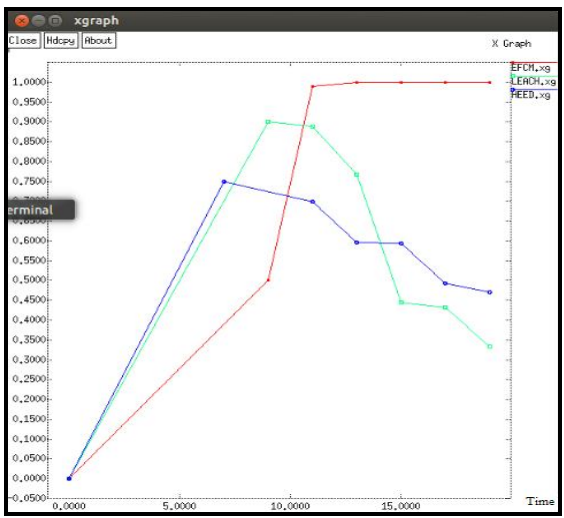

Fig. 9. Packet Delivery ratio vs. pause time.

Fig. 10 specifies the energy possessed by a sensor node at a given point of time. Since the energy is the main constraint in wireless sensor network so the energy

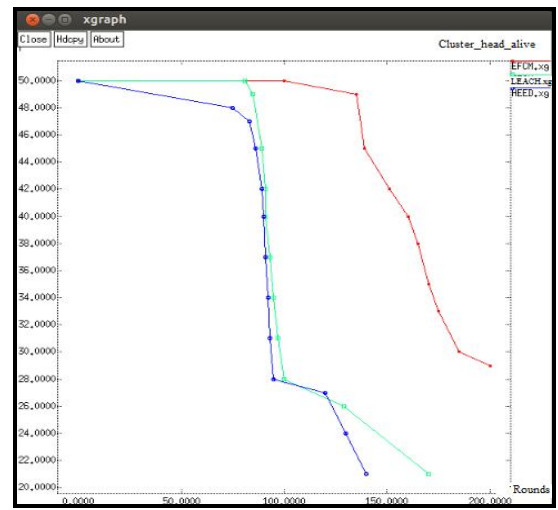

Fig. 11. Cluster-head failure.

## VI. CONCLUSION

In this paper we have proposed and implemented EFCM protocol algorithm. This algorithm performs better in variable sized networks as compared with the state of art methods. Our algorithm adopts a non-homogeneous.

However our algorithm results in some overhead due to context switching of cluster head. However this is overshadowed by high coverage fair distribution of load and thereby enhancing the lifetime of a network.